\documentclass[10pt]{iopart}

\usepackage{graphicx}

\begin{document}

\letter{}

\title[Response of the Heavy-Fermion Superconductor CeCoIn$_5$ to Pressure]
{Response of the Heavy-Fermion Superconductor CeCoIn$_5$ to Pressure:
Roles of Dimensionality and Proximity to a Quantum-Critical Point}

\author{M. Nicklas$^{\rm 1}$, R. Borth$^{\rm 2}$, E. Lengyel$^{\rm 2}$, P. G. Pagliuso$^{\rm 1}$, J. L.
Sarrao$^{\rm 1}$, V. A. Sidorov$^{\rm 1,}$\footnote{Permanent
address: Institute for High Pressure Physics, Russian Academy of
Science, Troitsk.}, G. Sparn$^{\rm 2}$, F. Steglich$^{\rm 2}$,
and J. D. Thompson$^{\rm 1}$}

\address{$^{\rm 1}$ Los Alamos National Laboratory Los Alamos, NM 87545, USA \\ $^{\rm 2}$
Max-Planck-Institute for the Chemical Physics of Solids, D-01187
Dresden, Germany}

\begin{abstract}
We report measurements of the pressure-dependent superconducting
transition temperature $T_c$ and electrical resistivity of the
heavy-fermion compound CeCoIn$_5$. Pressure moves CeCoIn$_5$ away
from its proximity to a quantum-critical point at atmospheric
pressure. Experimental results are qualitatively consistent with
theoretical predictions for strong-coupled, d-wave
superconductivity in an anisotropic 3D superconductor.

\end{abstract}


\nosections to appear in: {\it J. Phys.: Condens. Matter {\bf13}
No 44 (5 November 2001) L905-L912}

\nosections Although heavy-fermion superconductivity has been
known for over two decades, a microscopic theory of this
broken-symmetry state has remained elusive. Experiments have
established, however, that the unconventional superconductivity in
heavy-fermion compounds most likely is mediated by magnetic
fluctuations, which also are responsible for enhancing the
effective mass of itinerant quasiparticles by two to three orders
of magnitude.\cite{1,2,3} In spite of 20 years of searching,
there has been, until very recently, only one example of a
Ce-based heavy-fermion compound, CeCu$_2$Si$_2$, that is
superconducting at atmospheric pressure.\cite{4} In contrast,
several examples of Ce-based heavy-fermion antiferromagnets have
been found in which superconductivity appears as their N{\'e}el
temperature is tuned toward $T=0$ with the application of
pressure.\cite{5,6,7,8} This recent series of discoveries
re-enforces the belief not only that a magnetic interaction is
responsible for superconductivity in these materials but also
that heavy-fermion superconductivity may be favored at a
particular 'soft-spot' in phase space, i.e., near a
quantum-critical point.\cite{9}

In heavy-fermion superconductors, the characteristic
spin-fluctuation temperature $T_{sf}$ appears to set the scale
for $T_c$, in much the same way that the characteristic phonon
frequency does in the BCS theory of conventional superconductors.
The most direct measure of $T_{sf}$ comes from the linewidth of
neutron quasi-elastic scattering, but it frequently is estimated
from the magnitude of the specific heat Sommerfeld coefficient
$\gamma\propto 1/T_{sf}$. A very large Sommerfeld coefficient, the
hallmark of heavy-fermion behavior, implies, then, a small
$T_{sf}$ and consequently low $T_{c}$. As might be expected, the
$T_{c}$'s of all pressure-induced Ce superconductors are less
than 0.5~K and that of CeCu$_2$Si$_2$ is about 0.7~K. At the other
extreme are the cuprates, in which $T_{sf}$ and $T_{c}$ are both
larger by roughly two orders of magnitude than in heavy-fermion
systems. Besides $T_{sf}$, Monthoux and Lonzarich have argued,
specifically in the context of cuprates, that dimensionality is a
factor in determining the transition temperature of
magnetically-mediated superconductors.\cite{10} Their solution of
Eliashberg equations shows that, for all other factors being the
same, the mean-field $T_{c}$ of a 2-dimensional system will be
higher than its 3D analogue. There have been no examples,
however, of families of magnetically-mediated superconductors
that crystallize in related 3D and 2D structures to test these
suggestions.

The very recent observation of heavy-fermion superconductivity in
CeMIn$_5$ (M=Co, Ir, and Rh) offers the opportunity for at least a
qualitative inspection of the role of dimensionality. These
compounds form in the tetragonal HoCoGa$_5$ crystal structure that
can be viewed as layers of CeIn$_3$ and 'MIn$_2$' stacked
sequentially along their c-axis. CeCoIn$_5$\cite{11} and
CeIrIn$_5$\cite{12} superconduct at atmospheric pressure with bulk
$T_{c}$'s of 2.3 and 0.4~K, respectively, while
CeRhIn$_5$\cite{13} is an incommensurate antiferromagnet that
transforms to a superconductor with $T_{c}=2.1$~K at pressures
greater than about 1.6~GPa. For comparison, cubic CeIn$_3$, which
is the infinite-layer variant of these compounds, becomes a
superconductor at pressures near 2.5~GPa with a maximum $T_{c}$
of about 0.25~K.\cite{8} A very crude approximation of the role
of $T_{sf}$ in determining the magnitude of $T_{c}$ in these
materials comes from taking $T_c \propto T_{sf} \propto
1/\gamma$. For CeMIn$_5$, $T_{c}= 2.3$, 0.4 and 2.1~K for M=Co,
Ir and Rh, respectively. Their corresponding values of $\gamma$
are 250 \cite{11}, 750 \cite{12} and 380 \cite{Fisher} ${\rm
mJ/mol~K^2}$ for $T\geq T_c$. Within the CeMIn$_5$ family, then,
$T_c$ scales approximately with $1/\gamma \propto T_{sf}$. The
Sommerfeld coefficient of CeIn$_3$ at 2.5~GPa is not known, but
an upper limit should be its $P=0$ value of $130~{\rm
mJ/mol~K^2}$.\cite{Nasu} With all other factors assumed equal, we
would expect from this comparison that $T_c$ of CeIn$_3$ should
be over 3~K at 2.5~GPa, a factor of 10 higher than observed
experimentally. In view of the calculations by Monthoux and
Lonzarich, the high transition temperatures of CeMIn$_5$ members,
relative to that in CeIn$_3$, may be due in part to their layered
crystal structure.

In the following, we present results of pressure-dependent
measurements on CeCoIn$_5$ and discuss the relationship of its
layered structure particularly in the context of predictions by
Montoux and Lonzarich \cite{10}, and apparent proximity to a
quantum-critical point to an interpretation of its pressure
response.

Four-probe AC resistivity measurements, with current flow in the
tetragonal basal plane, were performed on a single crystal of
CeCoIn$_5$ grown from excess In flux. Pressure was produced in a
conventional Be-Cu clamp-type cell using pentene as the pressure
medium, and the clamped pressure at low temperatures was
determined from the shift in the inductively measured $T_{c}$ of
small piece of Pb located in close proximity to the sample. A
comparison of the $T_{c}$'s of Pb and CeCoIn$_5$ with literature
values at atmospheric pressure showed that there was a small
temperature gradient, at most 0.7\% the thermometer temperature,
between the sample volume and the calibrated Cernox thermometer
that was thermally anchored in the wall of the pressure cell.
Independent measurements of the pressure dependence of $T_{c}$ of
CeCoIn$_5$ were carried out over a more limited pressure range
using a small clamp cell in a Quantum Design Superconducting
Quantum Interference Device magnetometer, with Flourinert FC 75
(3M) as the pressure medium and a Pb manometer.

\begin{figure}
\centering
\includegraphics[angle=0,width=100mm,clip]{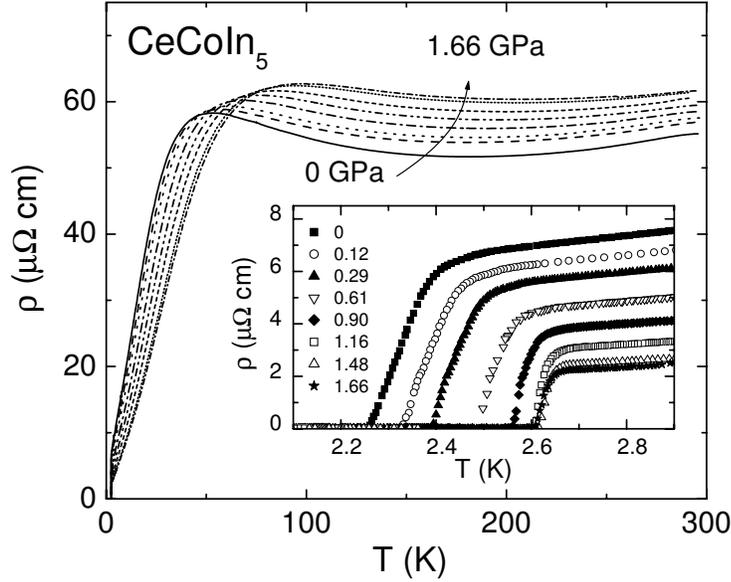}
\caption{\label{fig1} Temperature dependence of the in-plane
resistivity of CeCoIn$_5$ at various fixed pressures noted in the
inset. With increasing pressure, the resistivity increases at
high temperatures and decreases at low temperatures. The inset
shows the pressure response of the low-temperature resistivity
and superconducting transition temperature.}
\end{figure}

The pressure dependence of the resistivity of CeCoIn$_5$ over a
broad temperature range is plotted in figure~\ref{fig1}. At
atmospheric pressure, $\rho(T)$ reaches a maximum at $T_M=50$~K
that generally is associated with a crossover from incoherent
Kondo-like scattering at high temperatures to a heavy-fermion
band state at low temperatures. With increasing $P$, $T_M$ moves
to higher temperatures, and the resistivity above $T_M$ increases
monotonically; whereas, the resistivity below $T_M$ decreases.
This overall pressure response is typical of Ce-based
heavy-fermion materials and is consistent with a $P$-induced shift
of the characteristic spin-fluctuation temperature $T_{sf}$ and
spin-fluctuation scattering to higher temperatures.\cite{18} A
more detailed view of the low-temperature behavior, given in the
inset, shows that the resistivity just above $T_{c}$ decreases
from slightly over 0.06 to about 0.02~$\mu\Omega$m and $T_{c}$
increases from 2.25 to 2.6 K as the pressure is raised.

\begin{figure}
\centering
\includegraphics[angle=0,width=100mm,clip]{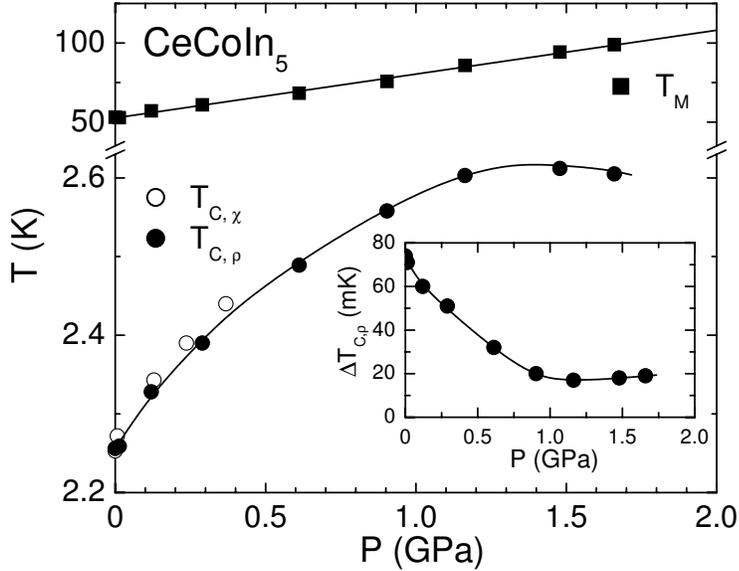}
\caption{\label{fig2}Pressure dependence of the temperature $T_M$
at which the resistivity is a maximum (solid squares) and the
superconducting transition temperature $T_{c}$, determined where
the resistivity is zero (solid circles) and by the mid-point of
the magnetic transition (open circles). The inset plots the
resistive transition width as a function of pressure. $\Delta
T_c$ was obtained from the temperatures at which the resistivity
reached 90\% and 10\% of its value just above onset of
superconductivity.}
\end{figure}

Explicit pressure dependencies of $T_M$ and $T_{c}$ are shown in
figure~\ref{fig2}. $T_M$ is linear in $P$ over the entire pressure
range, with ${\rm d}T_M/{\rm d}P=28$~K~GPa$^{-1}$. Assuming that
$T_M \propto T_{sf} \propto 1/\gamma$, such a large rate of
increase implies a rapid decrease in the electronic contribution
to the specific heat. Indeed, direct measurements of $\gamma(P)$
show \cite{19} this expected behavior, and we find that
$1/T_M(P)$ is linearly  proportional to the normal-state
$\gamma(P)$. The relative change $\partial\ln \gamma/\partial\ln
(1/T_M)= 1.2$ falls well within the range of values where this
comparison has been possible on other heavy-fermion
systems.\cite{18} Over this same pressure interval, $T_{c}$
increases initially at a rate ${\rm d}T_c/{\rm
d}P=0.5$~K~GPa$^{-1}$ and reaches a maximum value of 2.61~K near
1.5~GPa before decreasing. The inset shows that the resistive
transition width sharpens substantially from $\Delta T_c \approx
80$~mK at $P=0$ to around 20~mK for $P>0.9$~GPa.

It is interesting to compare these trends with the pressure
response of CeRhIn$_5$. In that case, pressure induces a rather
abrupt (as measured by the N{\'e}el temperature) change from
antiferromagnetic to superconducting states near 1.6~GPa. Just
beyond the critical pressure required for superconductivity, the
resistivity at $T\geq T_c$ is large and $\Delta T_c$ is relatively
broad, but with increasing $P$, $\rho(T\geq T_c)$ drops, $\Delta
T_c$ sharpens, ${\rm d}T_M/{\rm d}P$ is comparable to that found
in CeCoIn$_5$ \cite{13} and $T_{c}$ passes over a maximum near
3.0~GPa\cite{20}. From this comparison, it appears that
CeCoIn$_5$ at atmospheric pressure is very similar to CeRhIn$_5$
at a pressure of about 1.6~GPa, and one might infer that
CeCoIn$_5$ is near a quantum-critical point at atmospheric
pressure. Indeed, specific heat measurements on CeCoIn$_5$ in a
magnetic field sufficiently large to suppress $T_{c}$ to zero
find $C/T\propto-\ln T$ over more than a decade in temperature
above 100~mK, behavior that is accompanied by a resistivity
$\rho-\rho_0\propto T^{\alpha}$, with $\alpha \approx
0.95$.\cite{11,19,21} These temperature dependencies of $C/T$ and
$\rho$ are clear indications of a non-Fermi liquid state.\cite{22}

A variation in $T_{c}$ with tuning parameter, similar to $T_c(P)$
in figure~\ref{fig2}, also is predicted in the calculations of
Monthoux and Lonzarich \cite{10} for a p-wave or d-wave
superconductor in proximity to ferromagnetic and commensurate
antiferromagnetic instabilities, respectively. There is
substantial evidence pointing to the conclusion that CeCoIn$_5$
is a d-wave superconductor \cite{24,25} and, thus, we focus on
that case. Before comparing our measurements to these
calculations, we briefly summarize relevant assumptions and
results of those calculations. In ref.~\cite{10}, the authors
consider a square (2D) or cubic (3D) lattice in which the dominant
scattering of quasiparticles is magnetic, with a constant
spin-fluctuation temperature $T_{sf}$. From a mean-field solution
of Eliashberg equations, they obtain $T_{c}/T_{sf}$ as a function
of two dimensionless parameters, the square of the inverse
magnetic correlation length $\kappa^2$ and an effective
interaction parameter $g^2\chi_0/t$ arising from the exchange of
magnetic fluctuations. $\kappa^2\rightarrow 0$ corresponds to the
quantum-critical point where the magnetic correlation length
diverges. For fixed values of $\kappa^2>0$, $T_{c}/T_{sf}$
increases monotonically from zero as a function of $g^2\chi_0/t$
and saturates in the strong coupling limit, whereas, for fixed
$g^2\chi_0/t$, $T_{c}/T_{sf}$ decreases monotonically with
$\kappa^2$ for a 3D superconductor but passes over a maximum at
small $\kappa^2$ for the 2D case. In the absence of direct
measurements of either $\kappa^2$ or $g^2\chi_0/t$, we follow the
suggestions in ref.~\cite{10} and take pressure as a qualitative
measure of $\kappa^2$ and the resistivity $\rho_0$ just above
$T_{c}$ as a comparable measure of $g^2\chi_0/t$ for a comparison
of the theoretical calculations to experimental quantities.

\begin{figure}
\centering
\includegraphics[angle=0,width=100mm,clip]{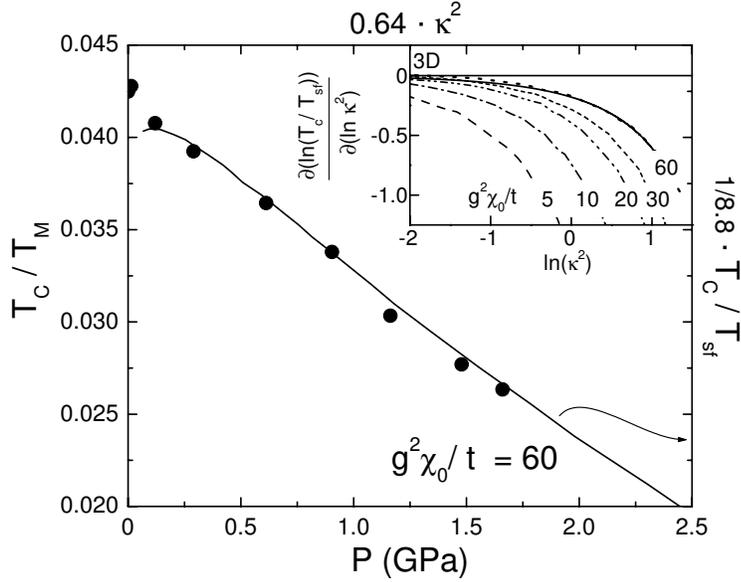}
\caption{\label{fig3} $T_c(P)/T_M(P)$, left ordinate, versus
pressure, bottom abscissa.  Experimental values plotted as solid
circles were taken from data in figure~\ref{fig2}. The solid line
is a theoretical prediction (right ordinate, top abscissa) of
$T_{c}/T_{sf}$ versus $\kappa^2$ for a 3D superconductor with
coupling constant $g^2\chi_0/t$=60. See text for details. In this
comparison, we have taken $0.64\kappa^2=P$ and $8.8 T_{sf}=T_M$. A
value of $g^2\chi_0/t$=60 was chosen for the comparison on the
basis of results shown in the inset, where theoretical values
(broken curves) of $\partial\ln(T_c/T_{sf})/\partial\ln(\kappa^2)$
versus $\ln \kappa^2$ for fixed $g^2\chi_0/t$ are plotted on the
same scale as experimental values (solid curve) of
$\partial\ln(T_c/T_M)/\partial\ln P$ versus $\ln P$.}
\end{figure}

The theoretical prediction of $T_{c}/T_{sf}$ versus $\kappa^2$
for a relatively strong-coupled, 2D, d-wave superconductor near an
antiferromagnetic quantum-critical point is qualitatively like
$T_c(P)$ for CeCoIn$_5$, particularly in the sense that both
exhibit a maximum. Because CeCoIn$_5$ is strong-coupled \cite{11}
and appears to be a d-wave superconductor near a quantum-critical
point, we might conclude from this comparison that it also is in
the 2D limit. The model calculations, however, take $T_{sf}$ to be
a rather large energy scale, a substantial fraction of the
tight-binding hopping integral, and thus to be independent of
$\kappa^2$ or pressure. Though this assumption may be quite
reasonable for cuprate superconductors, $T_{sf}$ in heavy-fermion
systems is typically much smaller, by one to two orders of
magnitude, than in the cuprates and pressure dependent.\cite{18}
Taking $T_M\propto T_{sf}$, we find that the ratio
$T_c(P)/T_M(P)$ is a monotonically decreasing function of
pressure, as shown in figure~\ref{fig3}. Explicitly including this
pressure dependence of $T_M$ makes a definitive comparison
between our experimental results and the theoretical expectation
for a 2D superconductor ambiguous.

\begin{figure}
\centering
\includegraphics[angle=0,width=90mm,clip]{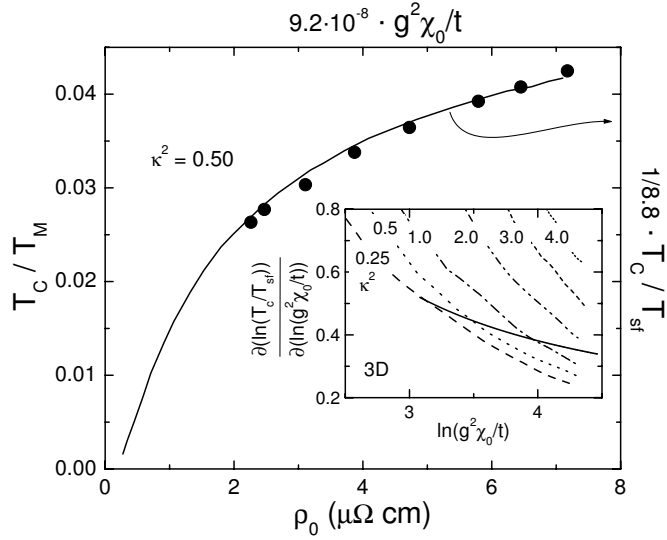}
\caption{\label{fig4} $T_c(P)/T_M(P)$, left ordinate, versus
electrical resistivity $\rho_0$ determined just above the onset of
superconductivity, bottom abscissa. Solid circles are
experimental data taken from figures~\ref{fig1} and~\ref{fig2}.
The solid line is a theoretical prediction (right ordinate, top
abscissa) of $T_{c}/T_{sf}$ versus $g^2\chi_0/t$ for a 3D
superconductor with $\kappa^2=0.5$. For the theoretical curve, we
have taken $9.2\times 10^{-8}g^2\chi_0/t =\rho_0$ and
$8.8T_{sf}=T_M$. The inset shows
$\partial\ln(T_c/T_{sf})/\partial\ln(g^2\chi_0/t)$ versus $\ln
(g^2\chi_0/t)$ for fixed $\kappa^2$ (broken curves) and the
experimentally determined $\partial\ln(T_c/T_M)/\partial\ln\rho_0$
versus $\ln\rho_0$ (solid curve) on the same scales.}
\end{figure}

\begin{figure}
\centering
\includegraphics[angle=0,width=90mm,clip]{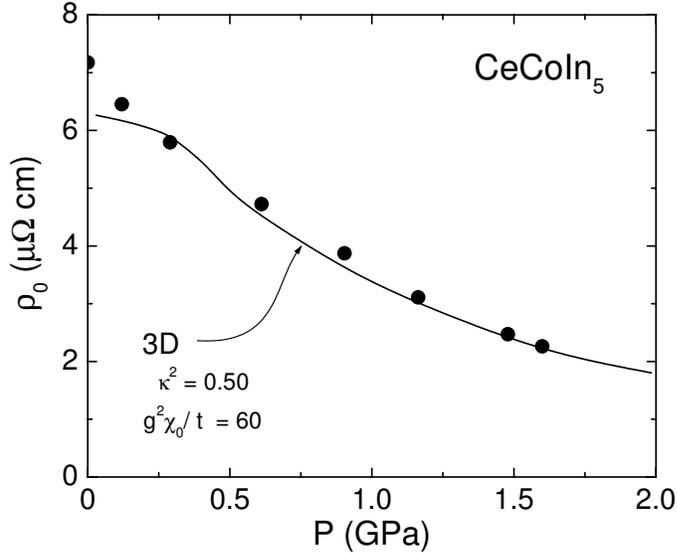}
\caption{\label{fig5} Resistivity $\rho_0$, determined just above
the onset of superconductivity, as a function of pressure. Solid
circles are experimental results and the solid curve is a
theoretical prediction for a 3D superconductor with
$\kappa^2=0.5$ and $g^2\chi_0/t=60$. For the theoretical curve,
we have used, as in figures~\ref{fig3} and~\ref{fig4},
$P=0.64\kappa^2$ and $\rho_0=9.2\times 10^{-8}g^2\chi_0/t$.}
\end{figure}

We emphasize this ambiguity by comparing theoretical predictions
for a 3D superconductor to $T_c(P)/T_M(P)$ versus pressure in
figure~\ref{fig3} and in figure~\ref{fig4} to $T_c/T_M$ as a
function of the electrical resistivity $\rho_0(P)$. We have
narrowed the phase space of possible theoretical parameters by
comparing logarithmic derivatives of the theoretical curves and
the experimental data. This type of comparison is independent of
proportionality factors between experimental observables and
theoretical parameters. As shown in the inset to
figure~\ref{fig3}, the experimental data
$\partial\ln(T_c/T_M)/\partial\ln P$ versus $\ln P$ are described
best by the theoretical prediction for
$\partial\ln(T_c/T_{sf})/\partial\ln\kappa^2$ versus
$\ln\kappa^2$ when the coupling constant $g^2\chi_0/t=60$. The
only free parameter in this comparison are a proportionality
factor between $T_{sf}$ and $T_M$ and a factor relating $P$ to
$\kappa^2$. We plot in the main body of figure~\ref{fig3} the
theoretical curve from ref.~\cite{10} for $T_c/T_{sf}$ as a
function of $\kappa^2$ at fixed $g^2\chi_0/t=60$. The value of
$T_c/T_{sf}$ has been multiplied by $1/8.8$ and $\kappa^2$ by
$0.64$. These scaling factors produce semiquantitative agreement
with the experimental data $T_c/T_M$ vs. $P$, shown by solid
circles. Likewise, in the inset of figure~\ref{fig4}, the
experimental determined $\partial\ln(T_c/T_M)/\partial\ln\rho_0$
versus $\ln\rho_0$ is compared to the theoretical predictions for
$\partial\ln(T_c/T_{sf})/\partial\ln(g^2\chi_0/t)$ versus
$\ln(g^2\chi_0/t)$. Best agreement between experiment and theory
is found for values of $\kappa^2$ less than $\sim1.0$. The main
body of figure~\ref{fig4} shows experimentally determined
$T_c/T_M$ versus $\rho_0$ and the theoretical prediction of
$T_c/T_{sf}$ as a function of $g^2\chi_0/t$ for fixed
$\kappa^2=0.5$. $T_c/T_{sf}$ has been multiplied by $1/8.8$, as
in figure~\ref{fig3}, and $g^2\chi_0/t$ has been multiplied by
$9.2\times10^{-8}$. Again, agreement between experiment and
theory is satisfactory. Overall, the theoretical predictions
shown in figures~\ref{fig3} and \ref{fig4} describe the
experimental results reasonably well. The value $\kappa^2=0.50$,
which provides a reasonable approximation of $T_c(P)/T_M(P)$
versus $\rho_0$, implies that CeCoIn$_5$ is close to a
quantum-critical point. Likewise, the theoretical curve for
$g^2\chi_0/t=60$, which corresponds to very strong coupling
\footnote{We note that RPA estimates of $g^2\chi_0/t$ given in
ref.~\cite{10} find expected values of the coupling parameter to
be in the range 5 to 20. The large jump in specific heat $\Delta
C/\gamma T_c \approx 4.5$ found for CeCoIn$_5$ \cite{12}
indicates very strong coupling that could exceed the RPA
estimate.}, qualitatively reproduces the $T_c(P)/T_M(P)$ versus
pressure data. (For simplicity in this comparison, in
figure~\ref{fig3} we have taken $\kappa^2=0$ at $P=0$, which
implies that CeCoIn$_5$ is precisely at a quantum-critical point
at atmospheric pressure. Somewhat better agreement between
experimental values and the theoretical prediction would result
if we had chosen $P\propto(\kappa^2_0+ \kappa^2)$, where
$\kappa^2_0$ corresponds to a large but finite magnetic
correlation length at $P=0$.) If these parameters have been
chosen correctly, they also should 'predict' $\rho_0(P)$ with no
other adjustments. From theoretical curves \cite{10} of
$T_c/T_{sf}$ versus $\kappa^2$, for fixed $g^2\chi_0/t=60$, and
of $T_c/T_{sf}$ versus $g^2\chi_0/t$, for fixed $\kappa^2=0.5$,
we can obtain $g^2\chi_0/t$ $(\propto \rho_0)$ as a function of
$\kappa^2$ $(\propto P)$, using $T_c/T_{sf}$ as the implicit
variable. The rather good agreement found in figure~\ref{fig5}
between direct measurements of $\rho_0(P)$ and the solid
theoretical curve for $g^2\chi_0/t$ versus $\kappa^2$ indicates
that our choice of $g^2\chi_0/t=60$ for varying $\kappa^2$ in
figure~\ref{fig3} and $\kappa^2=0.5$ for varying $g^2\chi_0/t$ in
figure~\ref{fig4} are reasonable. Strong coupling, d-wave
symmetry and proximity to a quantum-critical point are also
conclusions drawn in the preceding paragraph. Unlike the previous
conclusions, however, those drawn from solid curves in
figures~\ref{fig3}-\ref{fig5} are based on theoretical curves for
a 3D superconductor. Comparable agreement between experiment and
theory for the 2D limit was not possible.

The semi-quantitative agreement between experiment and theory
must be considered somewhat fortuitous. As noted earlier, the
calculations of ref.~\cite{10} assume that $T_{sf}$ is
independent of pressure, and, further, that the coupling constant
$g^2\chi_0/t$ and $\kappa^2$ are independent parameters. The
substantial pressure dependence of $T_M$ and specific heat
Sommerfeld coefficient in CeCoIn$_5$ point to an intrinsic
pressure dependence of its characteristic spin-fluctuation
temperature. Though we have used $\rho_0$ as an approximation to
$g^2\chi_0/t$, more direct measurements of the coupling, from
specific heat measurements on CeCoIn$_5$ as a function of
pressure \cite{19}, find that $\Delta C/\gamma T_c|_{T_c}$ is a
monotonically decreasing function of $P$, qualitatively similar
to $\rho_0(P)$. Therefore, it appears that $T_{c}$, $T_{sf}$,
$g^2\chi_0/t$ and $\kappa^2$ are intimately coupled in CeCoIn$_5$
in a much more complex way than assumed in ref.~\cite{10}. In
spite of these difficulties, the general trends found
experimentally and in the theoretical calculations are those that
might be expected from what is known about magnetically-mediated
superconductivity and CeCoIn$_5$ specifically, and we believe
that, at least qualitatively, the two are consistent. Further
refinement of theory to account for the interdependence of
parameters would allow a more straightforward comparison to
experiment. Beyond this is the issue of dimensionality that has
not been resolved. However, very recent calculations by Monthoux
and Lonzarich \cite{27} show that any amount of anisotropy
enhances $T_c$ above a strictly 3D value for any $\kappa^2$ and
$g^2\chi_0/t$. A comparison of CeCoIn$_5$ to cubic CeIn$_3$
supports this conclusion and suggests that CeCoIn$_5$ should be
considered to be an anisotropic 3D system.

In summary, a comparison to CeRhIn$_5$ indicates that the
electronic state of CeCoIn$_5$ at atmospheric pressure is
analogous to that of CeRhIn$_5$ at a pressure of about 1.6~GPa,
where superconductivity develops as its N{\'e}el transition
disappears. Thus, from its pressure response as well as its
properties at atmospheric pressure, CeCoIn$_5$ appears to be
close to a quantum-critical point. We have analyzed data in terms
of theoretical predictions for a d-wave superconductor in both 2D
and 3D limits and find that, though the 3D predictions
semi-quantitatively reproduce experimental observations, this
agreement must be taken with caution. The analysis, however, does
suggest avenues for extending the calculations to make a
comparison to experiment more meaningful. It is unlikely that
CeCoIn$_5$ should be considered a formally 2D system, but its
clear electronic \cite{15,16} and crystallographic anisotropy
plays a role in determining $T_c$ at both atmospheric and applied
pressures.

\nosections We thank P.~Monthoux for helpful discussions. Work at
Los Alamos was performed under the auspices of the US Department
of Energy. Work at the MPI-CPfS was conducted within the FERLIN
program of the ESF.

\newpage

\end{document}